\documentclass[a4paper,11pt]{article}
\pdfoutput=1
\usepackage{jheppub,MnSymbol}

\newcommand{\beq}{\begin{equation}}
\newcommand{\eeq}{\end{equation}}
\newcommand{\bea}{\begin{eqnarray}}
\newcommand{\eea}{\end{eqnarray}}

\preprint{MS-TP-19-29}

\title{Slepton pair production with aNNLO+NNLL precision}

\author[a]{Juri Fiaschi,}
\author[a]{Michael Klasen}
\author[a]{and Marthijn Sunder}

\affiliation[a]{Institut f\"ur Theoretische Physik, Westf\"alische Wilhelms-Universit\"at
 M\"unster, Wilhelm-Klemm-Stra\ss{}e 9, D-48149 M\"unster, Germany}

\emailAdd{fiaschi@uni-muenster.de}
\emailAdd{michael.klasen@uni-muenster.de}
\emailAdd{mpasunder@uni-muenster.de}

\abstract{We present a calculation of slepton pair production at the LHC
  at next-to-next-to-leading logarithmic (NNLL) accuracy, matched to approximate
  next-to-next-to-leading order (aNNLO) QCD corrections. We collect the relevant
  analytical formulae, discuss the matching of logarithmically enhanced
  and fixed-order results and describe the transformation of parton densities
  and hadronic cross sections to and from Mellin space. Numerically, we
  find a moderate increase of invariant-mass distributions and total cross
  sections with respect to our previous results at next-to-leading logarithmic (NLL)
  accuracy matched to next-to-leading order (NLO), and more importantly a further
  significant reduction of the
  factorisation and renormalisation scale dependence that stabilises our
  predictions to the permil level. The dependence on other supersymmetric
  parameters like squark and gluino masses and sbottom mixing that enter only
  at NLO is found to be weak, i.e.\ less than two percent, as expected.
}


\begin{document}
\maketitle
\flushbottom

\section{Introduction}
\label{sec:1}

The search for supersymmetric (SUSY) particles is an important current
research topic at CERN's Large Hadron Collider (LHC). The reason is that
SUSY is a well-motivated extension of the Standard Model (SM) of particle
physics that can solve a significant number of shortcomings of this model.
Important examples of SUSY solutions to SM problems are the stabilisation
of the Higgs boson mass and a possible candidate for dark matter, which
typically is the lightest neutralino, a mixture of the fermionic
partners of the neutral electroweak gauge and Higgs bosons
\cite{Nilles:1983ge,Haber:1984rc}.
Sleptons, the scalar partners of the SM leptons, are usually also among the
lightest SUSY particles \cite{AguilarSaavedra:2005pw}. While LHC searches
already constrain squarks and gluinos, the SUSY partners of quarks and
gluons, to the mass range above 1 or 2 TeV \cite{ATLAS:2019vcq,Sirunyan:2019ctn},
the limits on left-handed selectron and smuon masses are less stringent and
lie at 550 and 560 GeV, respectively \cite{Aad:2019vnb,Sirunyan:2018nwe}.
Staus can even be as light as 390 GeV \cite{ATLAS:2019ucg,Sirunyan:2019mlu}.

Experimental SUSY searches at the LHC rely on precise theoretical predictions
that go beyond leading order (LO) in perturbative QCD \cite{Dawson:1983fw,Bozzi:2004qq} and
include not only next-to-leading order (NLO) QCD \cite{Baer:1997nh}
and SUSY-QCD corrections \cite{Beenakker:1999xh}, but that also resum the
contributions that are logarithmically enhanced. These enhancements can otherwise
spoil the convergence of the perturbative series. They occur at small transverse
momenta of the produced slepton pair \cite{Bozzi:2006fw}, close to the production
threshold \cite{Bozzi:2007qr,Fuks:2013lya}, or both \cite{Bozzi:2007tea}. 
Threshold resummation corrections not only increase the production cross section,
thereby enhancing the discovery ranges or exclusion limits, but also reduce its
dependence on the unphysical factorisation and renormalisation scales and thus
render the theoretical predictions more accurate. Together with resummation-improved
parton density functions (PDFs) \cite{Bonvini:2015ira}, also the PDF uncertainty
can in principle be reduced \cite{Fiaschi:2018xdm,Fiaschi:2019etm},
even though in practice these
PDFs must currently be fitted to smaller data sets than global NLO analyses and
thus still have larger errors. Similar calculations have been performed for
gaugino and higgsino pairs \cite{Debove:2009ia,Debove:2010kf,Debove:2011xj,%
  Fuks:2012qx,Fuks:2014nha,Fiaschi:2018hgm}, gluinos and gauginos
\cite{Fuks:2016vdc,Fuks:2017bsw},
and additional gauge bosons \cite{Fuks:2007gk,Jezo:2014wra,Jezo:2015ida,Klasen:2016qux},
are available within the public code RESUMMINO \cite{Fuks:2013vua}
and are regularly employed in the experimental analyses by ATLAS
\cite{Aad:2019vvf} and CMS \cite{Sirunyan:2018iwl}. Predictions have also recently
been made for the high-luminosity (HL) and high-energy (HE) phases of the
LHC \cite{CidVidal:2018eel}.

In this paper, we take our precision calculations for slepton pair production
to the next level by resumming not only the leading and next-to-leading
logarithms (NLL), but also the next-to-next-to-leading logarithms (NNLL) and
matching them not only to the full NLO QCD and SUSY-QCD corrections, but also an
approximate next-to-next-to-leading order (aNNLO) calculation in QCD. The corresponding
analytical formulae are available in the literature \cite{Vogt:2000ci,%
Kidonakis:2003tx, Kidonakis:2007ww} and are collected here to make the paper
self-contained. Similar calculations, based on full NLO SUSY-QCD
calculations \cite{Beenakker:1996ch,Beenakker:1997ut}, have also been
performed previously for squarks and gluinos \cite{Beenakker:2014sma} and stops
\cite{Beenakker:2016gmf} and are available through
the public code NNLL-fast \cite{Beenakker:2016lwe}.

The paper is organised as follows: In Sec.\ \ref{sec:2}, we describe our
analytical approach and in particular how threshold logarithms can be
resummed at NNLL accuracy, matched to a fixed-order calculation up to
NNLO and how the PDFs and hadronic cross sections are transformed to and
from Mellin space. Our numerical results are contained in Sec.\ \ref{sec:3}.
This section starts with a discussion of the QCD and SUSY input parameters,
followed by a demonstration of how the NNLL and aNNLO contributions affect
the differential cross section in particular at high invariant masses.
We then show the effects of the new contributions on the total cross
section, its dependence on the factorisation and renormalisation scales
as well as on other SUSY parameters like the squark and gluino masses or
the trilinear coupling governing squark mixing in the bottom sector.
The ensuing conclusions are presented in Sec.\ \ref{sec:4}.

\section{Analytical approach}
\label{sec:2}

The hadronic invariant mass distribution for the production of slepton pairs,
\bea
 M^2{d\sigma_{AB}\over dM^2}(\tau)&=&\sum_{a,b}\int_0^1dx_adx_bdz
 \left[x_af_{a/A}(x_a,\mu_F^2)\right]
 \left[x_bf_{b/B}(x_b,\mu_F^2)\right]\nonumber\\
 &&\times\left[z\sigma_{ab}(z,M^2,\mu_R^2,\mu_F^2)\right]\delta(\tau-x_ax_bz),
 \label{eq:2.1}
\eea
is obtained from a convolution of the parton density functions (PDFs) $f_{a,b/A,B}$,
that depend on the longitudinal momentum fractions $x_{a,b}$ of the partons $a,b$
in the external hadrons $A,B$ and the factorisation scale $\mu_F$, with the partonic
cross section $\sigma_{ab}$, that depends on the squared invariant mass of the
produced sleptons $M^2$, its ratio $z=M^2/s$ (whereas $\tau=M^2/S$) to the partonic
(hadronic) center-of-mass energy $s$ ($S$), and the renormalisation and factorisation
scales $\mu_R$ and $\mu_F$, respectively.

While the leading order (LO) cross section \cite{Dawson:1983fw,Bozzi:2004qq}
and the virtual next-to-leading order (NLO) corrections are proportional to
$\delta(1-z)$ \cite{Baer:1997nh,Beenakker:1999xh}, the kinematic mismatch in the
cancellation of infrared divergences among the virtual and real corrections of
order $n$ introduces large logarithmic remainders proportional to
\bea
 \alpha_s^n(\mu_R^2) \left[{\ln^m(1-z)\over1-z}\right]_+ &,& {\rm where} \ m\leq2n-1,
 \label{eq:2.2}
\eea
which close to threshold ($z\to1$) spoil the convergence of the perturbative
series in $\alpha_s$ and therefore have to be resummed to all orders
\cite{Sterman:1986aj,Catani:1989ne}.
After performing a Mellin transformation,
\beq
 F(N)=\int_0^1dy\,y^{N-1}F(y),
\eeq
of the PDFs and partonic cross section in Eq.\ (\ref{eq:2.1}), the hadronic
cross section $\sigma_{AB}$ factorises, the singular terms in Eq.\ (\ref{eq:2.2})
turn into large logarithms of the Mellin variable $N$,
\bea
 \left[{\ln^m(1-z)\over1-z}\right]_+ & \to & \ln^{m+1}N+\dots,
\eea
and the partonic cross section $\sigma_{ab}$ can be written in the exponentiated form
\beq
 \sigma_{ab}^{\rm (res.)}(N,M^2,\mu_R^2,\mu_F^2) =
 H_{ab}(M^2,\mu_R^2,\mu_F^2)\exp[G_{ab}(N,M^2,\mu_R^2,\mu_F^2)]
 + \mathcal{O}\left(\frac{1}{N}\right).
 \label{eq:2.5}
\eeq
Here, the exponent $G_{ab}$ is universal and contains all the logarithmically
enhanced contributions in the Mellin variable $N$, while the hard function
$H_{ab}$ is independent of $N$, though process-dependent.

\subsection{Threshold resummation at NNLL accuracy}

Up to next-to-next-to-leading logarithmic (NNLL) accuracy, the exponent
$G_{ab}$ can be written as
\beq
 G_{ab}(N,M^2,\mu_R^2,\mu_F^2) =
 L G_{ab}^{(1)}(\lambda) +
 G_{ab}^{(2)}(\lambda, M^2, \mu_R^2,\mu_F^2) +
 \alpha_s G_{ab}^{(3)}(\lambda, M^2, \mu_R^2, \mu_F^2),
\eeq
where $\lambda = \alpha_s b_0 L$ and $L = \ln\bar{N} = \ln(Ne^{\gamma_E})$.
The coefficients of the QCD $\beta$-function are denoted by $b_n=\beta_n/(2\pi)^{n+1}$,
and the first three coefficients are given by \cite{Tarasov:1980au,Larin:1993tp}
\bea
 b_0 &=& \frac{1}{12 \pi} (11 C_A - 2 n_f), \\
 b_1 &=& \frac{1}{24 \pi^2} (17 C_A^2 - 5 C_A n_f - 3 C_F n_f), \\
 b_2 &=& \frac{1}{64 \pi^3} \left(\frac{2857}{54} C_A^3 - \frac{1415}{54} C_A^2 n_f
   - \frac{205}{18} C_A C_F n_f + C_F^2 n_f + \frac{79}{54} C_A n_f^2
   + \frac{11}{9} C_F n_f^2 \right)
\eea
with $C_A=N_C=3$, $C_F=(N^2-1)/(2N_C)=4/3$ and the number of active quark
flavours $n_f=5$. For Drell-Yan-like processes such as slepton or gaugino pair production
initiated by quarks and antiquarks only, the coefficients $G^{(i)}_{ab}=g_a^{(i)}+
g_b^{(i)}$ with $a=b=q$ can, e.g., be found up to next-to-leading logarithmic (NLL)
accuracy in Refs.~\cite{Bozzi:2007qr, Debove:2010kf}. At NNLL, one also needs
\cite{Vogt:2000ci} 
\bea
 g_q^{(3)}(\lambda) &=& \frac{A^{(1)} b_1^2}{2 \pi b_0^4} \frac{1}{1 - 2 \lambda} \left[2 \lambda^2 + 2 \lambda \ln(1-2\lambda) + \frac{1}{2}\ln^2(1-2\lambda)\right] \nonumber\\
&+& \frac{A^{(1)} b_2}{2 \pi b_0^3} \left[2 \lambda + \ln(1-2\lambda) + \frac{2\lambda^2}{1 - 2 \lambda}\right] +\frac{2A^{(1)}}{\pi} \zeta_2 \frac{\lambda}{1 - 2 \lambda} \nonumber\\
&-&  \frac{A^{(2)} b_1}{(2 \pi)^2 b_0^3} \frac{1}{1 - 2 \lambda} \left[2 \lambda^2 + 2 \lambda + \ln(1-2\lambda)\right] +\frac{A^{(3)}}{\pi^3 b_0^2} \frac{\lambda^2}{1 - 2 \lambda} - \frac{D^{(2)}}{2 \pi^2 b_0} \frac{\lambda}{1 - 2 \lambda}  \nonumber\\
&+&  \frac{A^{(1)} b_1}{2 \pi b_0^2} \frac{1}{1 - 2 \lambda} \left[2 \lambda + \ln(1-2\lambda)\right]\ln\left(\frac{M^2}{\mu_R^2}\right) +\frac{A^{(1)}}{2 \pi} \left[\frac{\lambda}{1 - 2 \lambda}\ln^2\left(\frac{M^2}{\mu_R^2}\right) - \lambda\ln^2\left(\frac{\mu_F^2}{\mu_R^2}\right)\right]\nonumber\\
&-& \frac{A^{(2)}}{2 \pi^2 b_0} \left[\frac{\lambda}{1 - 2 \lambda}\ln\left(\frac{M^2}{\mu_R^2}\right) - \lambda\ln\left(\frac{\mu_F^2}{\mu_R^2}\right)\right].
\eea
Here, the universal process-independent coefficients are given by
\cite{Moch:2005ba}
\bea
 A^{(1)} &=& 2 C_F, \\
 A^{(2)} &=& 2 C_F \left[C_A \left(\frac{67}{18} - \zeta_2 \right) - \frac{5}{9} n_f\right], \\
 A^{(3)} &=& \frac{1}{2} C_F \Big[C_A^2 \left(\frac{245}{24} - \frac{67}{9}\zeta_2 + \frac{11}{6}\zeta_3 + \frac{11}{5}\zeta_2^2 \right) +C_F n_f \left(2\zeta_3 - \frac{55}{24} \right) \nonumber \\
 &+& C_A n_f \left(\frac{10}{9}\zeta_2 - \frac{7}{3} \zeta_3 - \frac{209}{108} \right) - \frac{n_f^2}{27} \Big]
\eea
and \cite{Vogt:2000ci} 
\beq
 D^{(2)}=2C_F\left[C_A\left( -{101\over27}+{11\over3}\zeta_2+{7\over2}\zeta_3\right)
 +n_f\left({14\over27}-{2\over3}\zeta_2\right)\right].
\eeq

\subsection{Hard matching coefficients up to NNLO}

The hard $N$-independent part of the Mellin-transformed partonic cross section
in Eq.\ (\ref{eq:2.5}),
\beq
 H_{ab} (M^2,\mu_R^2,\mu_F^2) =
 \sigma^{(0)}_{ab} \mathcal{C}_{ab}(M^2,\mu_R^2,\mu_F^2),
\eeq
can be perturbatively expanded in terms of the Mellin-transformed LO cross
section $\sigma^{(0)}_{ab}$ and
\beq
 \mathcal{C}_{ab}(M^2,\mu_R^2,\mu_F^2) =
 \sum_{n=0}\left(\frac{\alpha_s}{2\pi}\right)^n \mathcal{C}_{ab}^{(n)}(M^2,\mu_R^2,\mu_F^2),
\eeq
where the hard matching coefficients
\beq
 \mathcal{C}_{ab}^{(n)}(M^2,\mu_R^2,\mu_F^2) =
 \left(\frac{2\pi}{\alpha_s}\right)^n \left[\frac{\sigma^{(n)}_{ab}}{\sigma^{(0)}_{ab}}\right]_{\rm N-ind.}
\eeq
are obtained from the finite ($N$-independent) terms in the ratio of the
$n$-th order cross section over the LO one. The coefficients up to next-to-next-to-leading
order (NNLO) can be obtained from Refs.\ \cite{Kidonakis:2003tx, Kidonakis:2007ww} and are
given by
\bea
 \mathcal{C}_{ab}^{(0)} &=&1,\\
 \mathcal{C}_{ab}^{(1)} &=& C_F\left[\frac{4}{3}(\pi^2 - 6) - 3 \log\left(\frac{\mu_F^2}{M^2}\right)\right], \\
 \mathcal{C}_{ab}^{(2)} &=& \frac{C_F}{720} \bigg\{ 5(-4605 C_A + 4599 C_F + 762 n_f) + 20 \pi^2 (188 C_A - 297 C_F - 32 n_f) \\ \nonumber
 &-& 92 \pi^4 (C_A - 6 C_F) + 180 (11 C_A + 18 C_F - 2 n_f)\log^2\left(\frac{\mu_F^2}{M^2}\right) \\ \nonumber
 &-& 160 (11 C_A - 2 n_f)(6 - \pi^2)\log\left(\frac{\mu_R^2}{M^2}\right) + 80 (151 C_A - 135 C_F + 2 n_f) \zeta_3 \\ \nonumber
 &+& 20 \log\left(\frac{\mu_F^2}{M^2}\right) \bigg[-51 C_A + 837 C_F + 6 n_f - 4 \pi^2 (11 C_A + 27 C_F - 2 n_f) \\ \nonumber
 &+& (-198 C_A + 36 n_f)\log\left(\frac{\mu_R^2}{M^2}\right) + 216 (C_A - 2 C_F)\zeta_3\bigg]  \bigg\}.
\eea
By including the coefficients up to NNLO, the resummation of logarithmically enhanced
contributions is improved, since also beyond NNLO in $\alpha_s$ the finite terms
are multiplied by threshold logarithms.

\subsection{Fixed-order matching and inverse Mellin transform}

Although near to threshold the resummed cross section is a valid approximation,
outside this region the normal perturbative calculation should be used. A reliable
prediction in all kinematic regions is then obtained through a consistent
matching of the two results with
\beq
 \sigma_{ab} = \sigma_{ab}^{\rm (res.)} + \sigma_{ab}^{\rm (f.o.)} - \sigma_{ab}^{\rm (exp.)}\,.
\eeq
Here, the resummed cross section $\sigma_{ab}^{\rm (res.)}$ in Eq.~\eqref{eq:2.5}
has been re-expanded to NNLO, yielding $\sigma_{ab}^{\rm (exp.)}$, and subtracted
from the fixed-order calculation $\sigma_{ab}^{\rm (f.o.)}$ in order to avoid the
double counting of the logarithmically enhanced contributions. At ${\cal O}
(\alpha_s^2)$ we then obtain
\bea
 \sigma^{\rm (exp.)}_{ab}(N,M^2\!\!&\!,\!&\!\mu_R^2,\mu_F^2) =
 \sigma^{(0)}_{ab} \mathcal{C}_{ab}(M^2,\mu_R^2,\mu_F^2) \exp[G_{ab}(N,M^2,\mu_R^2,\mu_F^2)] \nonumber \\
 &=& \sigma^{(0)}_{ab} \left[1 + \left(\frac{\alpha_s}{2\pi}\right) \mathcal{C}_{ab}^{(1)} + \left(\frac{\alpha_s}{2\pi}\right)^2 \mathcal{C}_{ab}^{(2)} +\dots\right]\left[1 + \left(\frac{\alpha_s}{2\pi}\right) \mathcal{K}^{(1)} + \left(\frac{\alpha_s}{2\pi}\right)^2 \mathcal{K}^{(2)} +\dots\right] \nonumber \\
 &=& \sigma^{(0)}_{ab}\left[1 + \left(\frac{\alpha_s}{2\pi}\right) \left(\mathcal{C}_{ab}^{(1)} + \mathcal{K}^{(1)}\right) + \left(\frac{\alpha_s}{2\pi}\right)^2 \left(\mathcal{C}_{ab}^{(2)} + \mathcal{K}^{(2)} + \mathcal{C}_{ab}^{(1)}\mathcal{K}^{(1)}\right) + \dots\right].
\eea
The coefficients of the expanded exponential term can be organised in powers of $L$ as
\bea
 \mathcal{K}^{(1)} &=& \mathcal{K}^{(1,1)} L + \mathcal{K}^{(1,2)} L^2, \\
 \mathcal{K}^{(2)} &=& \mathcal{K}^{(2,1)} L + \mathcal{K}^{(2,2)} L^2 + \mathcal{K}^{(2,3)} L^3 + \mathcal{K}^{(2,4)} L^4.
\eea
Explicitly, they are given by \cite{Kidonakis:2003tx, Kidonakis:2007ww}
\bea
 \mathcal{K}^{(1,1)} &=& 4 C_F \log\left(\frac{\mu_F^2}{s}\right), \\
 \mathcal{K}^{(1,2)} &=& 4 C_F, \\
 \mathcal{K}^{(2,1)} &=& -\frac{C_F}{27} \bigg\{56 n_f - 404 C_A + 3\log\left(\frac{\mu_F^2}{s}\right) \bigg[20n_f + 2 C_A (-67 + 3\pi^2) \nonumber\\
 &+& 3 (11C_A - 2 n_f) \left(\log\left(\frac{\mu_F^2}{\mu_R^2}\right) - \log\left(\frac{\mu_R^2}{s}\right)\right)\bigg] + 378 C_A \zeta_3 \bigg\}, \\
 \mathcal{K}^{(2,2)} &=& \frac{2}{9} C_F \bigg[-10 n_f + 67 C_A - 3 C_A \pi^2 + 36 C_F \log^2\left(\frac{\mu_F^2}{s}\right) \nonumber \\
 &+& (33 C_A - 6 n_f) \log\left(\frac{\mu_R^2}{s}\right) \bigg], \\
 \mathcal{K}^{(2,3)} &=& \frac{4}{9} C_F \left[11 C_A - 2 n_f + 36 C_F \log\left(\frac{\mu_F^2}{s}\right)\right], \\
 \mathcal{K}^{(2,4)} &=& 8 C_F^2.
\eea
The SUSY-QCD (squark-gluino loop) corrections are only matched at NLO,
since they are not known beyond this order \cite{Beenakker:1999xh}.
In this sense, our results are accurate to approximate NNLO (aNNLO) plus
NNLL precision. This approximation is justified by the fact that the
SUSY-QCD corrections are subdominant due to the large squark and gluino
masses.

Having computed the resummed and the perturbatively expanded results
in Mellin space, we must multiply them with the $N$-moments of the PDFs and
perform an inverse Mellin transform,
\bea
 \label{eq:2.51}
 M^2{d\sigma_{AB}\over d M^2}(\tau)&=&{1\over2\pi i}\int_{{\cal C}_N} d N 
 \tau^{-N} M^2{d\sigma_{AB}(N)\over d M^2},
\eea
in order to obtain the hadronic cross section as a
function of $\tau=M^2/S$. Special attention must be paid to the
singularities in the resummed exponents $G_{ab}^{(1,2,3)}$, which are situated at
$\lambda=1/2$ and are related to the Landau pole of the perturbative coupling
$\alpha_s$. In order to avoid this pole as well as those in the Mellin moments of the PDFs
related to the small-$x$ (Regge) singularity $f_{a/A}(x,\mu_0^2)\propto x^\alpha
(1-x)^\beta$ with $\alpha<0$, we choose an integration contour ${\cal C}_N$
according to the {\em principal value} procedure proposed in
Ref.~\cite{Contopanagos:1993yq} and the {\em minimal prescription} proposed in
Ref.~\cite{Catani:1996yz}. We define two branches,
\bea
  {\cal C}_N:~~ N=C+ze^{\pm i\phi}~~{\rm with}~~ z\in[0,\infty[,
  \label{eq:IT:Nbra}
\eea
where the constant $C$ is chosen such that the singularities of the $N$-moments
of the PDFs lie to the left and the Landau pole to the right of the
integration contour. Formally, the angle $\phi$ can be chosen in the range
$[\pi/2,\pi[$, but the integral converges faster if $\phi>\pi/2$.
The Mellin moments of the PDFs are obtained by fitting to the parameterisations
tabulated in $x$-space the functional form used by the MSTW collaboration \cite{Martin:2009iq}
\beq
 f(x) = A_0\, x^{A_1}\,  (1 - x)^{A_2} \, \left(1 + A_3 \, \sqrt{x} + A_4 \, x + A_5 \, x^{\frac{3}{2}}\right) + A_6\, x^2 + A_7\, x^{\frac{5}{2}} \,,
\eeq
which has the advantage that it can be transformed analytically with the result
\bea
F(x) &=& A_{0} \, \Gamma\left(y\right) \, \mathrm{B'}\left(A_1 + N, y\right) + A_3 \, \mathrm{B'}\left(A_1 + N + \frac{1}{2}, y\right) + A_4 \mathrm{B'}\left(A_1 + N + 1, y\right)\nonumber \\ 
&+& A_5 \mathrm{B'}\left(A_1 + N + \frac{3}{2}, y\right) + A_6\mathrm{B'}\left(A_1 + N + 2, y\right) + A_7 \mathrm{B'}\left(A_1 + N + \frac{5}{2}, y\right) \,.
\eea
Here, $y = A_2 + 1$ and $\mathrm{B'}(x,y) = \mathrm{B}(x,y)/\Gamma(y) = \Gamma(x)/\Gamma(x + y)$.
We verified that we obtain good fits not only for the MMHT2014NLO118 \cite{Martin:2009iq},
but also for the CT14NLO fits \cite{Dulat:2015mca} up to large values of $x$ 
and for all typical factorisation scales, 
even though the latter are obtained with an ansatz that includes an exponential function.

\section{Numerical results for slepton pair production}
\label{sec:3}

In this section, we present numerical results for slepton pair production
at the LHC with aNNLO+NNLL
precision. We first discuss our choice of input parameters and demonstrate the
impact of threshold resummation on the invariant-mass distributions, after which
we show and discuss our experimentally more relevant predictions for the total
cross sections as a function of the slepton mass and other, subdominant SUSY
parameters.

\subsection{Input parameters}

Our numerical results for proton-proton collisions at LHC Run 2 with a
center-of-mass energy $\sqrt{S}$ of 13 TeV have been obtained with CT14 PDFs
\cite{Dulat:2015mca}, which we employ consistently at LO and NLO with the
corresponding partonic cross sections. While the PDF uncertainty in resummation
calculations can in principle be reduced by using also resummation-improved PDFs
\cite{Fiaschi:2018xdm,Fiaschi:2019etm},
the latter are fitted to a substantially smaller data
set than those at NLO, which unfortunately currently still results in a larger
PDF uncertainty \cite{Bonvini:2015ira}. We therefore use here NLO PDFs
with NLL and NNLL partonic cross sections and refer to Refs.\
\cite{Fiaschi:2018xdm,Fiaschi:2019etm} for a detailed discussion of PDF
uncertainties.
Since top (s)quarks do not enter
our calculations, all other five quark flavours are treated as massless, and
the QCD scale parameter $\Lambda$ is fixed accordingly to its CT14 values.
For our central predictions, the renormalisation and factorisation scales are
identified with the slepton mass. For scale uncertainty estimates, we employ
the seven-point method, i.e.\ the scales are varied individually by relative
factors of two, but not four.

Based on an integrated LHC luminosity of 139 (35.9) fb$^{-1}$ and for
sufficiently large mass differences with the lightest neutralino, the ATLAS
(CMS) collaboration has recently excluded left-handed selectrons below 550
(400) GeV. For two (not three, as stated in the ATLAS abstract and conclusion)
generations of mass-degenerate sleptons, the limit increases to 700 (450) GeV
\cite{Aad:2019vnb,Sirunyan:2018nwe}. We therefore adopt for the invariant-mass
distributions a default slepton mass of 1 TeV and use 700 GeV as the lower
mass limit for the total cross sections. Squarks and gluinos enter only at NLO
in virtual loop diagrams, and therefore their masses play only a subdominant role.
We adopt a squark and gluino mass of 1.3 TeV as our default value, which is
still allowed for not too large mass differences with the lightest neutralino,
even though the most stringent ATLAS (CMS) mass limits already reach 1.94
(1.63) and 2.35 (2.31) TeV, respectively \cite{ATLAS:2019vcq,Sirunyan:2019ctn}.
We will study the dependence on these parameters up to 2.5 TeV and see
that the dependence is indeed weak, as is the dependence on the mixing angle
in the case of bottom squarks.

\subsection{Invariant-mass distributions}

\begin{figure}
\begin{center}
\includegraphics[width=0.8\textwidth]{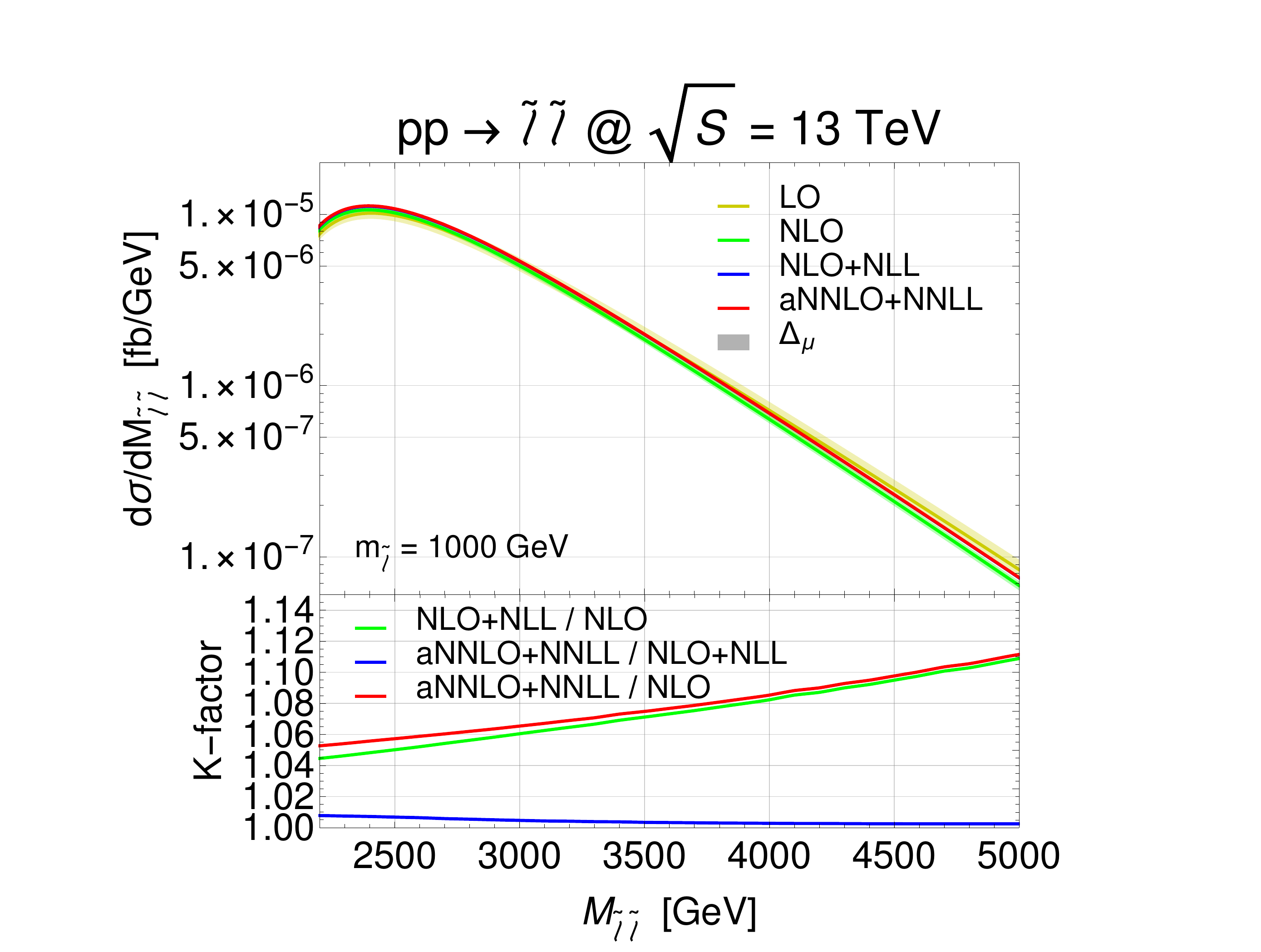}
\caption{Top: Invariant-mass distribution for left-handed selectron (or smuon)
  pair production at the LHC with a center-of-mass energy of $\sqrt{S}=13$ TeV
  for a fixed slepton mass of 1 TeV. Shown are results at LO (yellow), NLO
  (green), NLO+NLL (blue) and aNNLO+NNLL (red) together with the corresponding
  scale uncertainties (shaded bands).
  Bottom: Ratios ($K$ factors) of NLO+NLL over NLO (green), aNNLO+NNLL over
  NLO+NLL (blue) and aNNLO+NNLL over NLO (red) differential cross sections as
  a function of the invariant mass of the slepton pair.}
\label{fig:1}
\end{center}
\end{figure}

In Fig.\ \ref{fig:1} (top) we plot the invariant-mass distributions for
slepton pair production at the LHC with a center-of-mass energy of
$\sqrt{S}=13$ TeV and with LO (yellow), NLO (green), NLO+NLL (blue) and
aNNLO+NNLL (red) precision together with the corresponding scale uncertainties
(shaded bands). Since we do not take into account decays or detector
acceptances, these results are valid for both left-handed selectrons
and smuons of 1 TeV mass, while the cross sections for maximally mixed
staus or right-handed selectrons and smuons are typically smaller by
about a factor of 2 to 2.5 \cite{Fiaschi:2018xdm}. The cross section
rises with the third power of the slepton velocity and peaks at an
invariant mass that is considerably above the minimal value $2m_{\tilde{\ell}}$
before falling steeply off due to the $s$-channel propagator and the parton
luminosity \cite{Bozzi:2007qr}.

The effect of the higher-order corrections is best seen in Fig.\ \ref{fig:1}
(bottom) as ratios ($K$ factors) of NLO+NLL over NLO (green), aNNLO+NNLL over
NLO+NLL (blue) and aNNLO+NNLL over NLO (red) differential cross sections.
Resummation effects at NLL (green) accuracy become more important with respect
to the fixed (NLO) order as the invariant mass of the slepton pair approaches
the production threshold. The corresponding $K$ factor increases in the
invariant mass range of 2.2 to 5 TeV from 4.5\% to 11\%. The increase from
NLO+NLL to aNNLO+NNLL is much smaller as expected for a converging expansion,
and most visible at low invariant masses, where the constant terms at aNNLO
induce an increase by about 1\%.

\begin{figure}
\begin{center}
\includegraphics[width=0.8\textwidth]{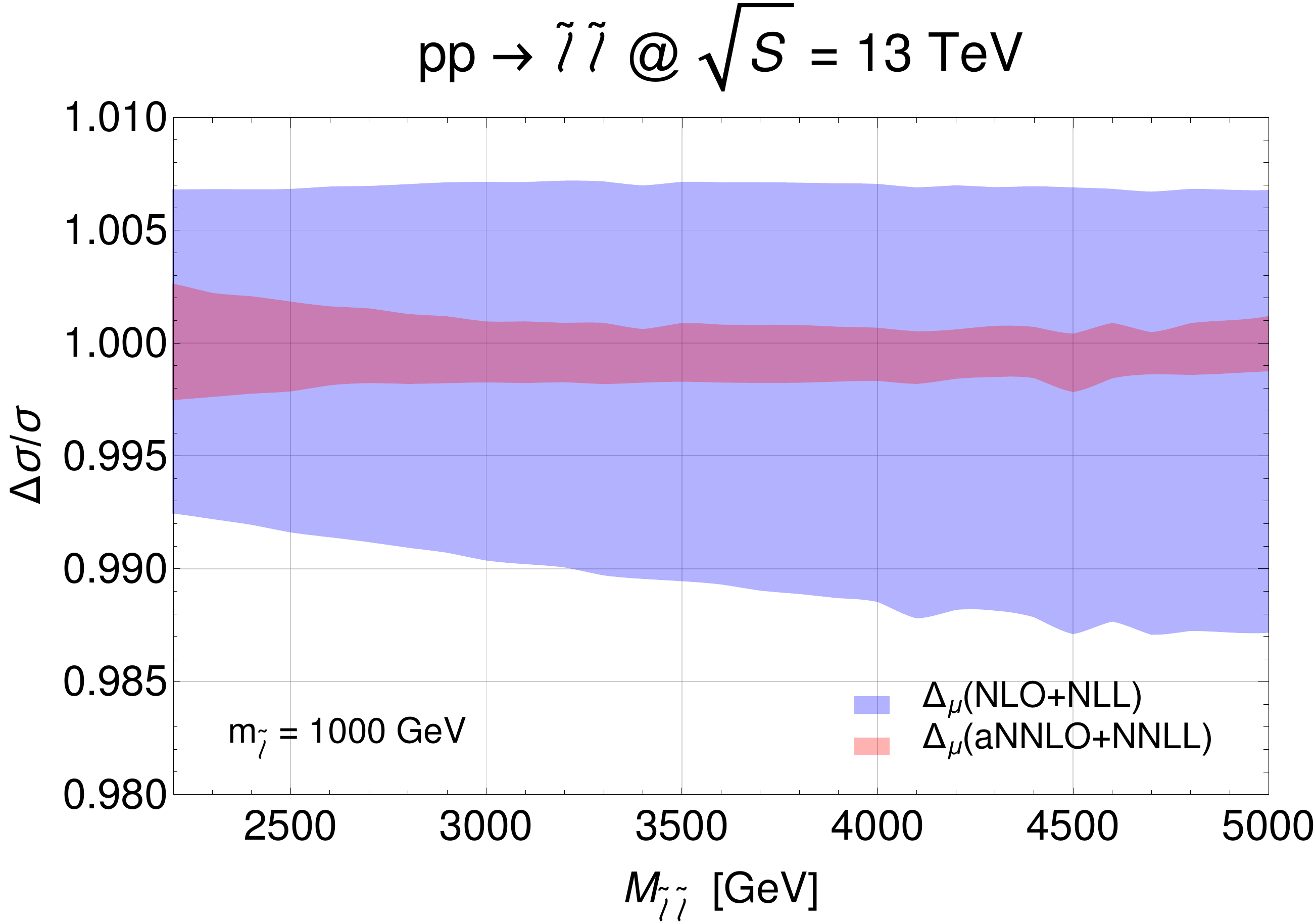}
\caption{Scale uncertainty of the invariant-mass distribution for
  left-handed selectron (or smuon) pair production at the LHC with a
  center-of-mass energy of $\sqrt{S}=13$ TeV for a fixed slepton mass
  of 1 TeV. Shown are the results at NLO+NLL (blue) and
  aNNLO+NNLL (red shaded band).}
\label{fig:2}
\end{center}
\end{figure}

Apart from the increase in cross section, which enhances the discovery range
for new particles at the LHC, a second important effect of resummation
calculations is the reduction in the theoretical uncertainty. It is
estimated by varying the renormalisation and factorisation scales following
the seven-point method. The
result for the invariant mass distribution is shown in Fig.\ \ref{fig:2}.
While the uncertainty remained already mostly below one percent and exceeded
this value very close to threshold at NLO+NLL (blue), the new contributions
at aNNLO+NNLL (red shaded band) reduce the uncertainty considerably further
to about one permil. Only at low invariant mass, i.e. far from threshold,
the uncertainty rises to about two permil. This demonstrates the
excellent stability of the expansion.

\subsection{Total cross sections}

\begin{figure}
\begin{center}
\includegraphics[width=0.8\textwidth]{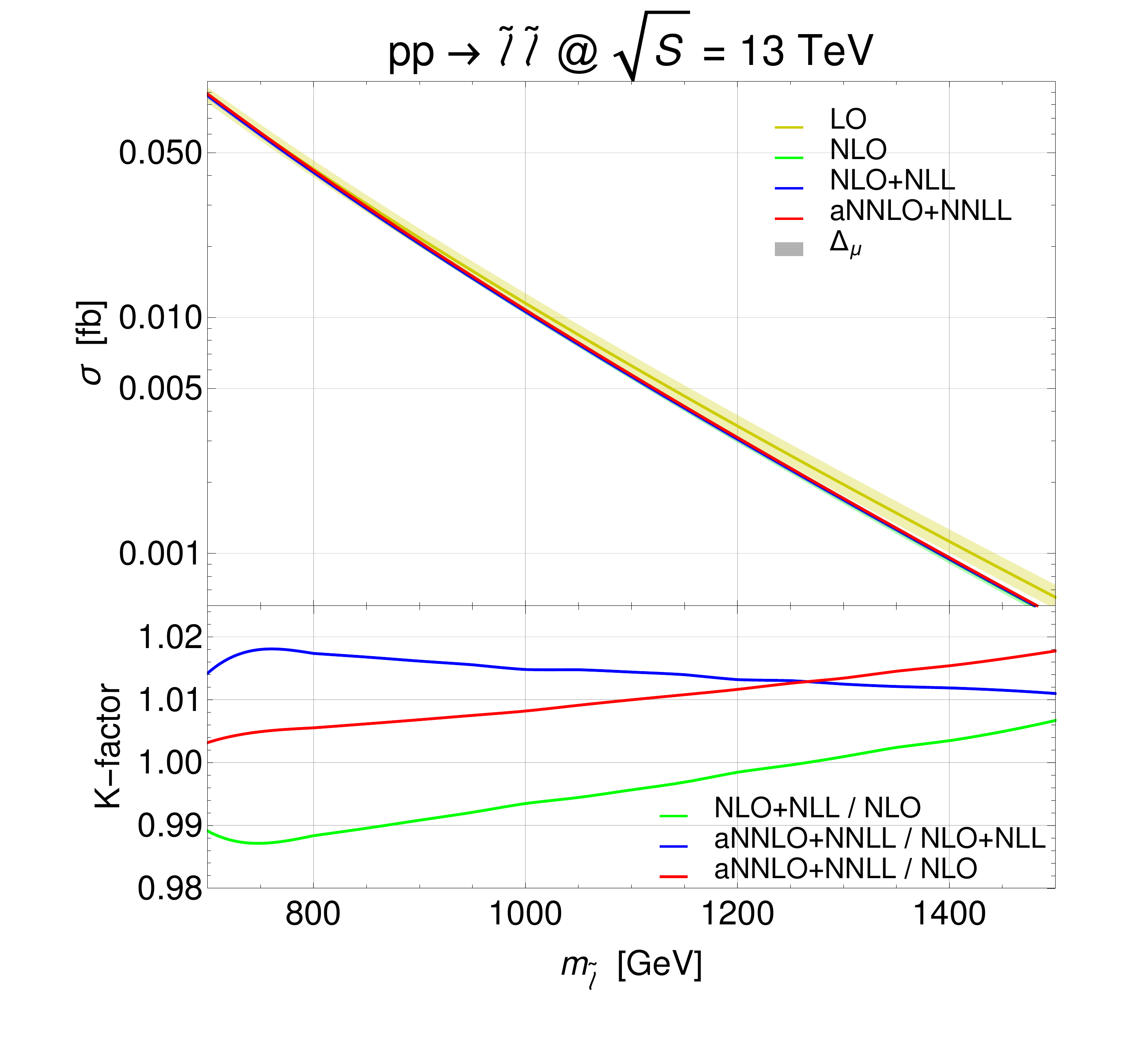}
\caption{Top: Total cross section for left-handed selectron (or smuon)
  pair production at the LHC with a center-of-mass energy of $\sqrt{S}=13$ TeV
  as a function of the slepton mass. Shown are results at LO (yellow), NLO
  (green), NLO+NLL (blue) and aNNLO+NNLL (red) together with the corresponding
  scale uncertainties (shaded bands).
  Bottom: Ratios ($K$ factors) of NLO+NLL over NLO (green), aNNLO+NNLL over
  NLO+NLL (blue) and aNNLO+NNLL over NLO (red) total cross sections as a
  function of the slepton mass.}
\label{fig:3}
\end{center}
\end{figure}

We now turn to our predictions for total cross sections for slepton pair
production at the LHC, which are directly applicable to determine experimental
discovery ranges or exclusion limits. To this end, we plot in Fig.\ \ref{fig:3}
(top) the total production cross section for left-handed selectron (or
smuon) pairs at the LHC with a center-of-mass energy of $\sqrt{S}=13$
TeV as a function of the slepton mass in the range 700 GeV to 1500 GeV.
In this range, the cross section falls from almost 0.1 fb to below 1 ab,
corresponding to more than 10 events at 700 GeV with the currently analysed
integrated luminosity of 139 fb$^{-1}$ to 3 events at 1 TeV with the LHC
Run 3 goal of 300 fb$^{-1}$ and a few events at 1.5 TeV with the high-luminosity
(HL) LHC goal of 3 ab$^{-1}$. The reduction of the scale uncertainty is visible
as a decrease in width of the predictions from LO (yellow shaded band) to the
higher orders (other colours).

The $K$ factors in Fig.\ \ref{fig:3} (bottom) show that the logarithmic
terms at NLL (green) and NNLL (red) first reduce, then enhance the cross
section by a few percent with respect to the NLO prediction as the slepton
mass increases. The aNNLO(+NNLL) terms lead in addition to an almost
constant increase over the NLO(+NLL) prediction of about one percent (blue).

\begin{figure}
\begin{center}
\includegraphics[width=0.8\textwidth]{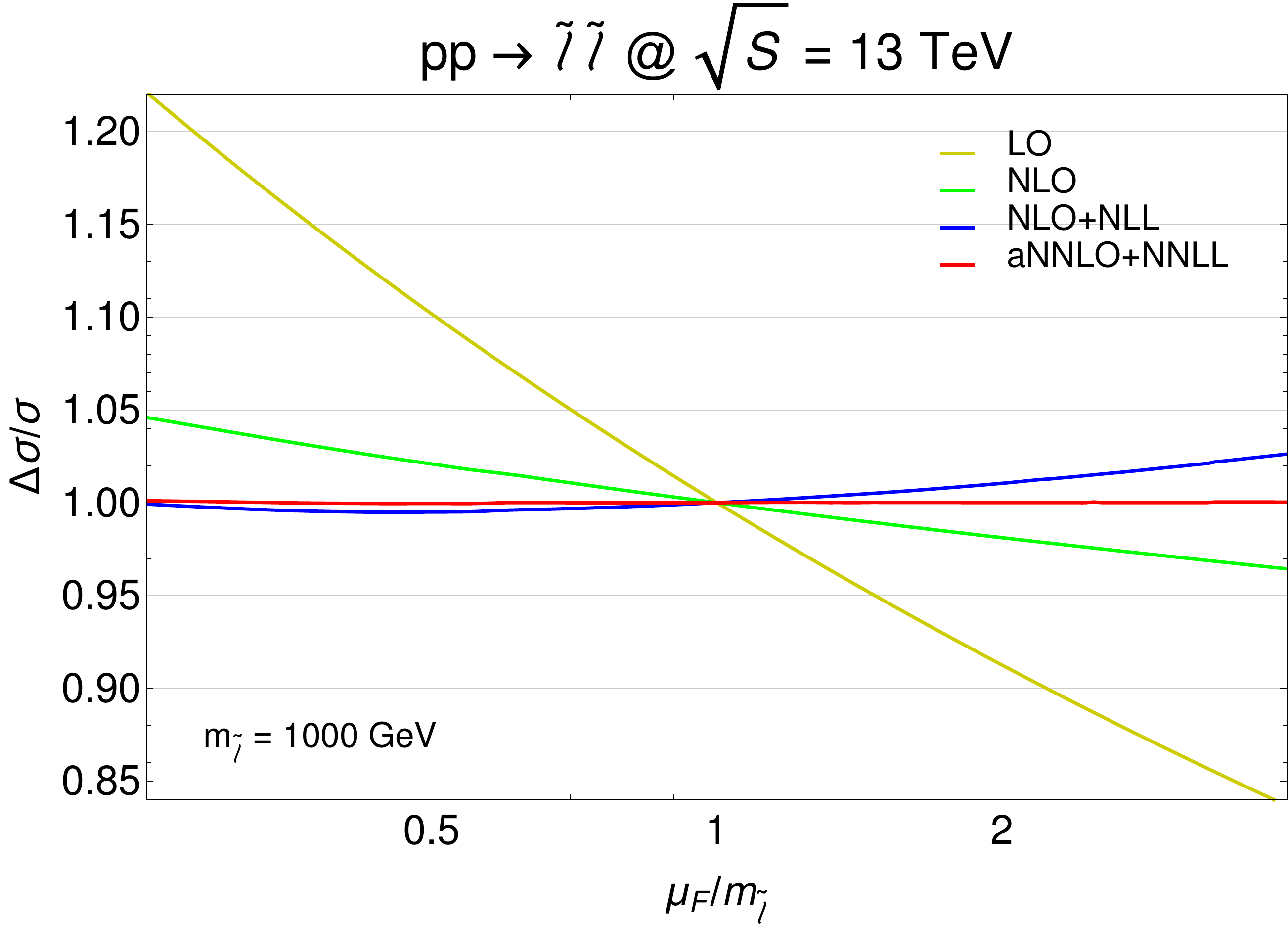}
\includegraphics[width=0.8\textwidth]{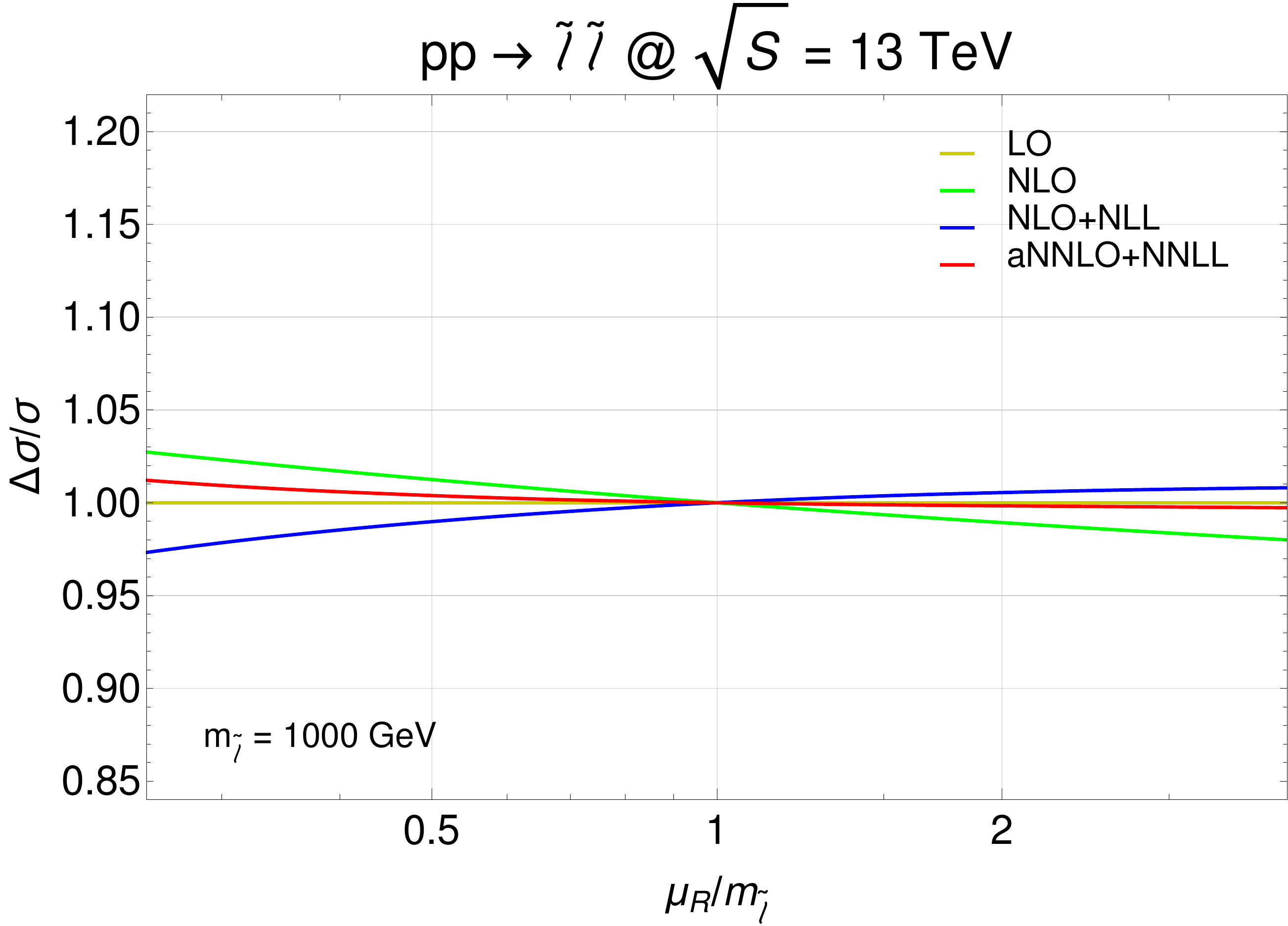}
\caption{Relative variation of the total cross section for slepton pair
  production as function of the factorisation (top) and renormalisation
  scale (bottom). Shown are results at LO (yellow), NLO (green), NLO+NLL
  (blue) and aNNLO+NNLL (red).}
\label{fig:4}
\end{center}
\end{figure}

As for the invariant mass distribution, it is important to study the
scale dependence at different levels of precision also for the total cross
section. The variation of the total slepton pair production cross section
at the LHC with 13 TeV center-of-mass energy with the factorisation scale
is shown in Fig.\ \ref{fig:4} (top), normalised to the cross section at the
central scale (the slepton mass of 1 TeV). The renormalisation scale is here
fixed to this value. While we observe a steeply falling dependence from the
PDFs at LO (yellow), it is already partially compensated at NLO through the
factorisation of collinear divergences (green), further reduced and somewhat
overcompensated at NLO+NLL (blue) and completely flat at aNNLO+NNLL (red).

Fig.\ \ref{fig:4} (bottom) shows the corresponding renormalisation scale
dependence, where now the factorisation scale remains fixed. The dependence
is only introduced at NLO, where $\alpha_s(\mu_R)$ falls with increasing
scale (green), since the LO cross section is of electroweak origin (yellow).
One then observes an oscillating behavior at NLO+NLL (blue) and aNNLO+NNLL
(red) with a variation that is reduced from 5\% at NLO to 1\% at aNNLO+NNLL.
This demonstrates again the excellent stability of the calculation.

\begin{figure}
\begin{center}
\includegraphics[width=0.8\textwidth]{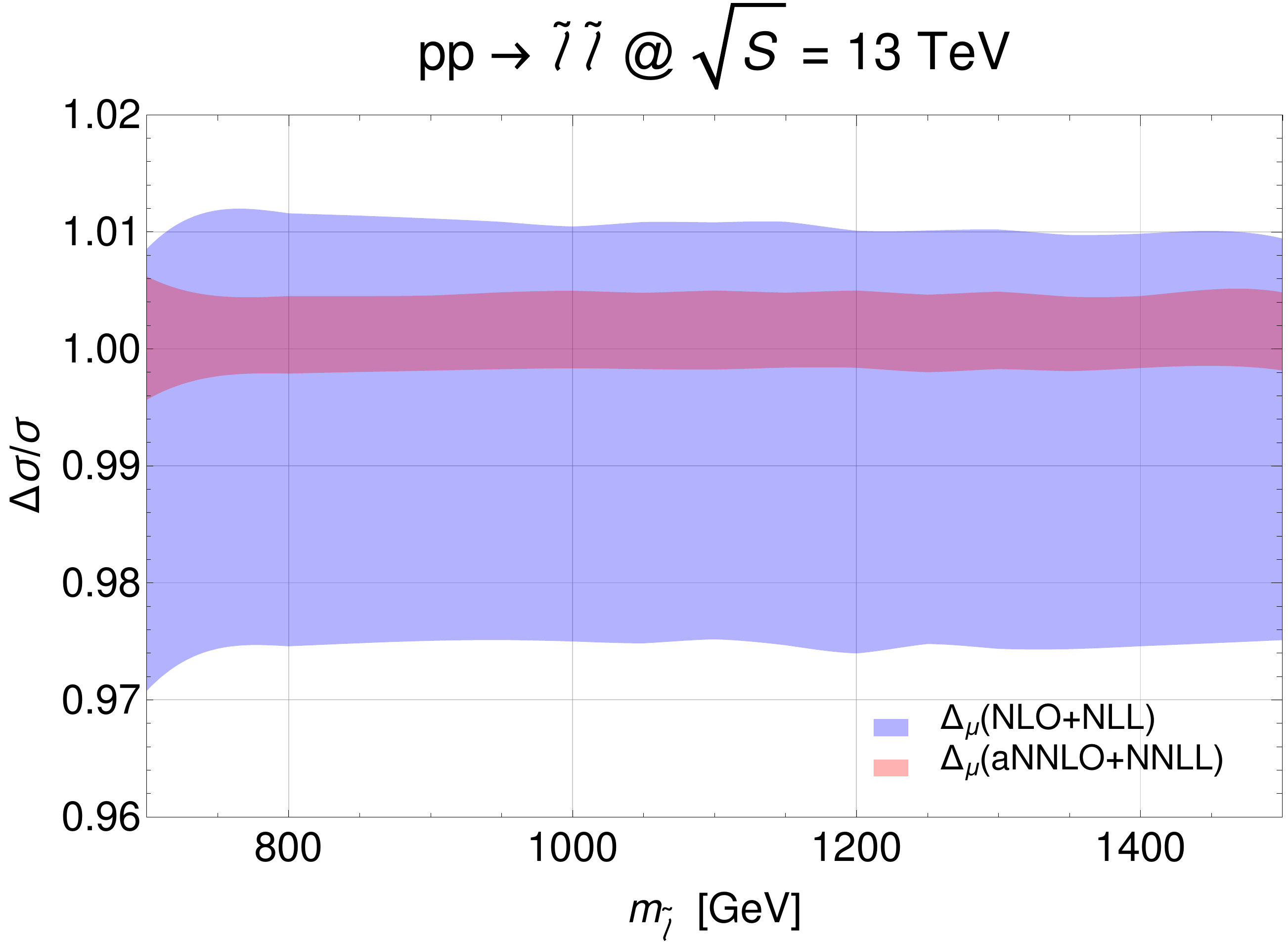}
\caption{Scale uncertainty of the total cross section for
  left-handed selectron (or smuon) pair production at the LHC with a
  center-of-mass energy of $\sqrt{S}=13$ TeV as a function of the slepton mass.
  Shown are the results at NLO+NLL (blue) and
  aNNLO+NNLL (red shaded band).}
\label{fig:5}
\end{center}
\end{figure}

The combined effect of varying the factorisation and renormalisation scales
with the seven-point method is shown in Fig.\ \ref{fig:5} as a function of
the slepton mass in the same range of 700 GeV to 1.5 TeV as considered above.
We observe an almost constant theoretical uncertainty of $-3$\% to $+1$\% at NLO+NLL (blue),
which is reduced to about $-0.2$\% to $+0.4$\% at aNNLO+NNLL (red shaded band)
and which is only slightly larger for small slepton masses.

\begin{figure}
\begin{center}
\includegraphics[width=0.8\textwidth]{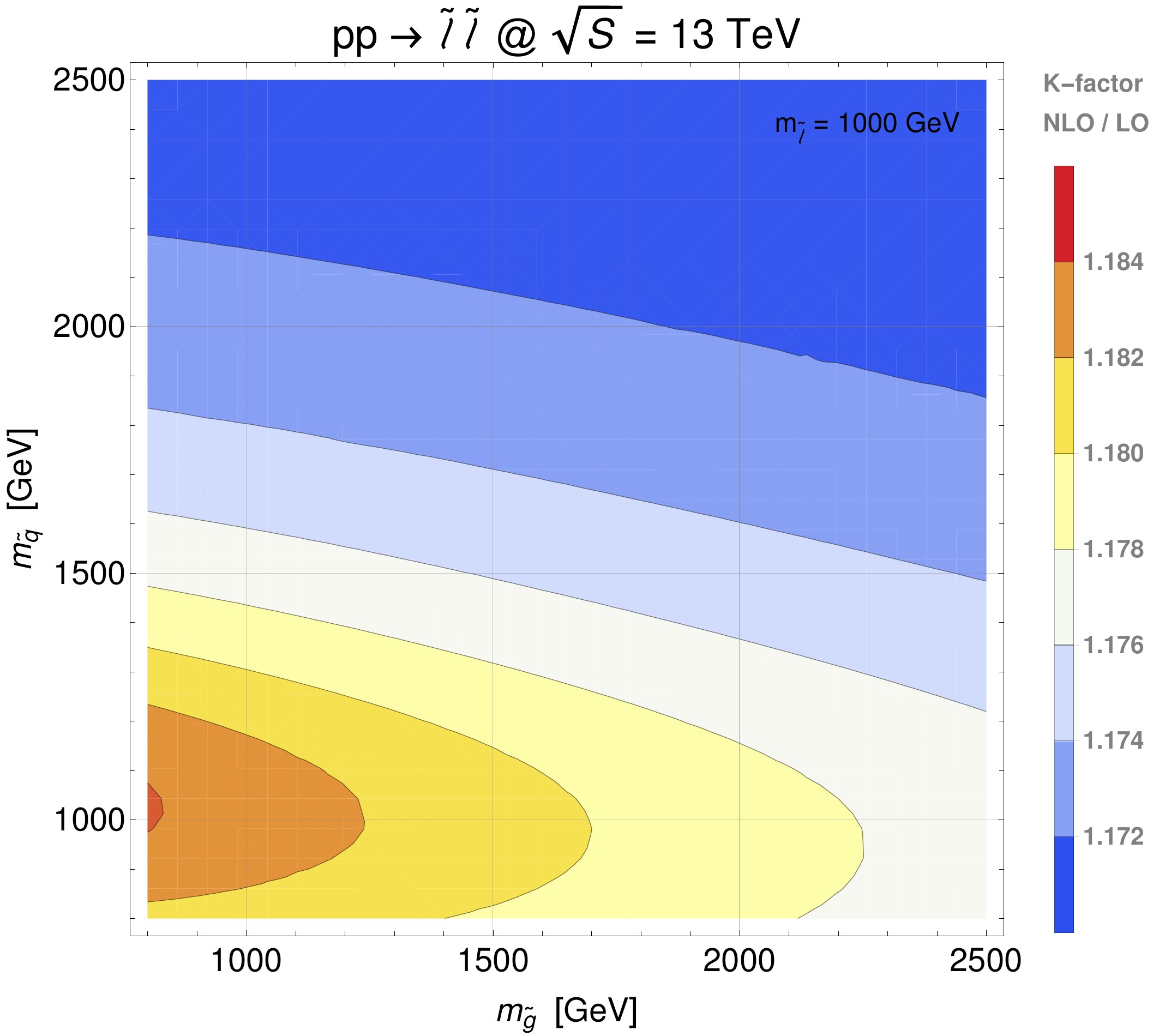}
\caption{Ratio ($K$ factor) of NLO over LO total cross sections (both
  with NLO PDFs) for left-handed selectron (or smuon) pair production
  at the LHC with a center-of-mass energy of $\sqrt{S}=13$ TeV as a
  function of the squark and gluino masses.}
\label{fig:6}
\end{center}
\end{figure}

The virtual corrections at NLO do not only introduce a dependence on the
renormalisation scale, but -- through the squarks and gluinos appearing in
the loops -- also a weak dependence on other SUSY masses. Resumming
logarithmically enhanced or adding approximate NNLO QCD, but not NNLO SUSY-QCD
contributions does not alter this dependence significantly. In Fig.\ \ref{fig:6}
we show the impact of other SUSY masses on the slepton pair production
cross section at the LHC as a colour-coded ratio of NLO over LO cross
sections in the squark-gluino mass plane. Overall, the dependence is
weak, as the $K$ factor varies only from 1.170 to 1.186, i.e.\ by less
than two percent. When the squark mass crosses the slepton mass at 1
TeV, the threshold behaviour in the triangle loop is clearly visible
and represents the dominant dependence. The gluino mass appears only
in the $t$-channel and is clearly less important. The squarks and
gluinos decouple and no longer influence the cross section, when their
masses reach the multi-TeV scale.

\begin{figure}
\begin{center}
\includegraphics[width=0.8\textwidth]{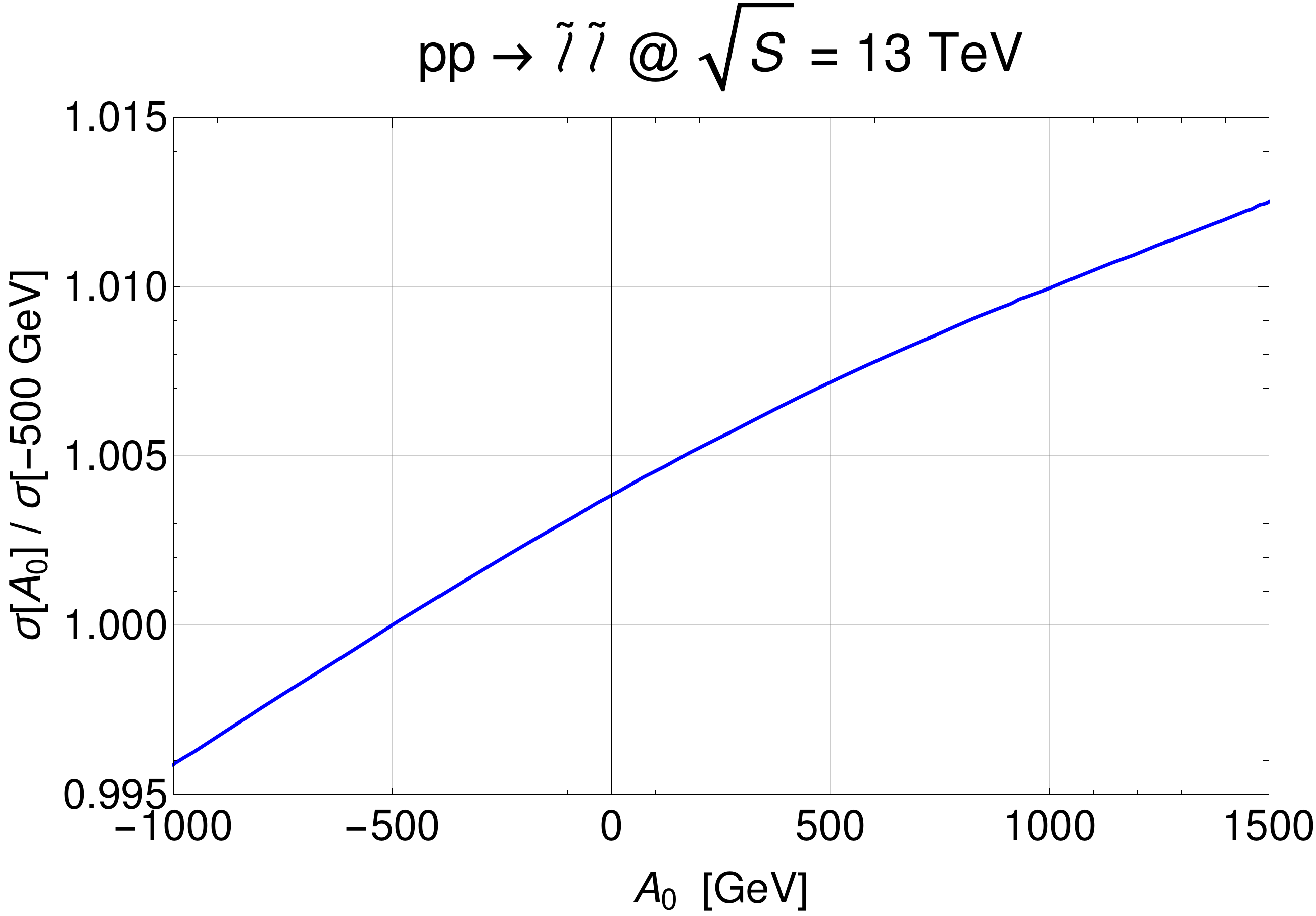}
\caption{Dependence of the NLO (or NLO+NLL or aNNLO+NNLL) total cross
  section on the common trilinear coupling $A_0$ that governs squark
  mixing in the sbottom sector. Shown is the ratio over the default
  scenario with $A_0=-500$ GeV.}
\label{fig:7}
\end{center}
\end{figure}

In simplified scenarios such as the phenomenological Minimal Supersymmetric
Standard Model (pMSSM) \cite{Berger:2008cq,Fuks:2017rio}
, it is common to assume a degeneracy of sfermion
masses. For the first two generations, it is then a good approximation to do
so also for the superpartners of the left- and right-handed fermions, since
the off-diagonal terms in the sfermion mass  matrix are proportional to
the corresponding fermion mass. This is different for the third generation,
where in the off-diagonal entries of the squark mass matrix the heavy-quark masses $m_t$
or $m_b$ multiply the combinations
\bea
 m_{LR}&=&A_0-\mu^\ast\left\{
 \begin{array}{l}
 \cot\beta\hspace*{3.mm}{\rm for~up-type~sfermions}
 \\ \tan\beta\hspace*{2.8mm}{\rm for~down-type~sfermions}
 \end{array}\right.\hspace*{3mm}
\eea
of the trilinear coupling $A_0$, the higgsino mass parameter $\mu$ and the ratio
of Higgs vacuum expectation values $\tan\beta=v_u/v_d$ for stops and
sbottoms, respectively. While stops do not enter our
calculations due to a negligible top quark PDF, sbottom mixing can
influence slepton pair production at NLO. This is demonstrated in
Fig.\ \ref{fig:7}, where we show the dependence of the total slepton
production cross section on the trilinear coupling $A_0$ normalised
to the cross section at the default value of $-500$ GeV. The other relevant SUSY parameters
have been set to $m_{LL}\simeq m_{RR}\simeq1.1$ TeV, $\mu\simeq0.8$ TeV
and $\tan \beta=40$. As expected, one observes an even weaker dependence
of the NLO cross section on the sbottom mixing than on the squark and
gluino masses, as it varies only from $-0.4$ to $+1.2$ percent.

\section{Conclusion}
\label{sec:4}

In conclusion, we have presented in this paper a calculation of threshold
resummation effects on slepton pair production at the LHC with NNLL accuracy
matched to approximate NNLO QCD corrections. We collected the relevant
analytical results from the literature and described the procedures, with
which we matched resummation and fixed-order results and performed the
transformation of PDFs and hadronic cross sections to and from Mellin space.
Numerically, we found only very moderate increases of invariant-mass
distributions and total cross sections with respect to our previous
calculations with NLO+NLL precision. More importantly, we observed very
significant reductions on the renormalisation and factorisation scale
dependences, that now stabilise our predictions to the permil level.
We also discussed briefly the dependence of the cross section on squark
and gluino masses that enter through virtual loop diagrams at NLO and
demonstrated that our calculations are also applicable to mixing
squarks, in particular of the third generation. Our results have been
implemented in the code RESUMMINO and will soon become available with
the next public release.

\section*{Acknowledgements}
\noindent
We thank B.\ Fuks for his collaboration on the off-diagonal squark loop
contributions and N.\ Kidonakis for useful discussions. This work has
been supported by the BMBF under contract 05H18PMCC1 and the DFG through
the Research Training Network 2149 ``Strong and weak interactions - from
hadrons to dark matter''.

\bibliographystyle{apsrev4-1}
\bibliography{bib}

\end{document}